\documentclass{aa}
\usepackage{graphicx}
\usepackage{epsfig}
\usepackage{rotating}
\usepackage{aalongtable,lscape}
\usepackage{txfonts}
\newcommand{\teff}{T$_{\rm eff}$}
\newcommand{\nli}{$\log$~n(Li)}

\newcommand{\vmi}{V$-$I}

\begin{document}
\title{Membership and lithium in the old, metal--poor open
cluster Berkeley~32
\thanks{Based on observations collected at ESO-VLT, Paranal Observatory,
Chile, Program number 74.D-0571(A)}}

%   \subtitle{}

   \author{S. Randich\inst{1} \and G. Pace\inst{1,2} 
         \and L. Pastori\inst{3} \and A. Bragaglia\inst{4}}

   \offprints{S. Randich, email:randich@arcetri.astro.it}

\institute{INAF/Osservatorio Astrofisico di Arcetri, Largo E. Fermi 5,
             I-50125 Firenze, Italy
\and
Centro de Astrofisica, Universidade de Porto, Rua das Estrellas, 
4150-762 Porto, Portugal
\and
INAF/Osservatorio Astronomico di Brera, Via E. Bianchi 46, I-23807, Italy
\and
INAF/Osservatorio Astronomico di Bologna, Via Ranzani 1, I-40127 Bologna, Italy}

\date{Received Date: Accepted Date}

% \abstract{}{}{}{}{}
% 5 {} token are mandatory

\abstract
{Measurements of lithium (Li) abundances in open clusters
provide a unique tool for following the evolution of this element
with age, metallicity, and stellar mass. 
In spite of the plethora of Li data
already available, the behavior of Li in solar--type stars has
so far been poorly understood.}
{Using FLAMES/Giraffe on the VLT, we obtained spectra of 157
candidate members of the old, metal--poor cluster Berkeley~32, 
to determine membership and to study the Li behavior
of confirmed members.
}
{Radial velocities were measured, allowing us to derive both the cluster
velocity and membership information for the sample stars. 
The Li abundances were obtained from the equivalent
width of the Li~{\sc i}~670.8~nm feature, using curves of growth.
}
{We obtained an average radial velocity of $105.2 \pm 0.86$~km/s, and
53 \% of the stars have a radial velocity consistent with
membership. The Li -- \teff~distribution of unevolved members
matches the upper envelope of M~67, as well as that of the
slightly older and more metal-rich NGC~188. No major 
dispersion in Li is detected.
When considering the Li distribution as a function of mass, however,
Be~32 members with solar-like temperature are less massive
and less Li-depleted than their counterparts in the other clusters.
The mean Li of stars 
in the temperature interval $5750 \leq \rm T_{\rm eff} \leq 6050$~K
is \nli=$2.47\pm 0.16$, less than a factor of two below the
average Li of the 600~Myr old Hyades,
and slightly above the average
of intermediate age (1--2~Gyr) clusters, the upper envelope of M67, and NGC~188.
This value is comparable to or slightly higher than
the plateau of Pop.~{\sc ii} stars. The similarity
of the average Li abundance of clusters of different age and metallicity,
along with its closeness to the halo dwarf plateau, is very intriguing
and suggests that,
whatever the initial Li abundance and the Li depletion histories,
old stars converge to almost the same final Li abundance.
}
{}


\titlerunning{Membership and lithium in Be 32}
\authorrunning{S. Randich et al.}
\maketitle
\section{Introduction}\label{intro}
Despite its low abundance (N(Li)/N(H) $\leq 2\times 10^{-9}$),
lithium (Li)
plays a fundamental role in different fields of astrophysics because of
how it is created and destroyed. The $^7$Li isotope is the heaviest element
produced during Big Bang nucleosynthesis
(BBN), and its primordial abundance strongly depends on $\eta$, the ratio
of baryons to photons (e.g., Steigman~\cite{steig_cast} and references therein).
At the same time, Li is easily depleted from stellar atmospheres:
$^7$Li is destroyed by proton reactions at
the relatively low temperature of $\sim 2.5$~MK, implying depletion with 
respect to the initial content
whenever a mixing process is present that
is able to transport surface material down to the deeper regions in
the stellar interior where this temperature is reached. 

Primordial Li abundance has been empirically estimated by means of
Li measurements in old Pop.~{\sc ii} stars and, theoretically, using
BBN calculations combined
with the baryon density derived from measurements of  the  cosmic
microwave background temperature fluctuations. The reliability of the
latter has dramatically  improved after data  from the WMAP  satellite
became available (Cyburt et al.~\cite{cyburt}; 
Spergel et al.~\cite{spergel}). Estimates from old
stars rely on the assumption that the hot (T$_{\rm eff}\geq$ 5800
K), metal poor, Pop.~{\sc ii} dwarfs do not deplete Li in their
atmosphere, hence represent the primordial
material from which they were born. 
The confidence on this assumption is based on the near
constancy of their Li abundance with metallicity, first discovered by
Spite \& Spite (\cite{spite82}) 
and generally referred to as the Spite plateau. 
Primordial Li abundance predicted from standard BBN calculations with
WMAP is a factor of two to three higher than that of the  Spite
plateau, which might indicate that halo stars have indeed undergone
some Li depletion, possibly due to microscopic diffusion,
as recently claimed by Korn et al. (\cite{korn},~\cite{korn1}). Only
through a full understanding of mixing inside stars and its dependence
on metallicity will we be able to reach a conclusion with complete
confidence. 

To achieve that, studies in Pop.~{\sc   i} stars,  and in
particular among open cluster members, are the most suitable observation
tool. Lithium (and beryllium) measurements in Galactic
open clusters (OCs) indeed allow us to empirically trace Li depletion with age, 
metallicity, and mass.
Focusing on solar-type stars, standard models of stellar
evolution (those including  convection only) predict that these stars  should
not suffer any Li depletion during the main sequence (MS), since their
convective zone does not reach the Li burning layer.  Surveys of Li in
a variety of OCs and, in particular, the comparison  of the Li
patterns of OCs of different age, have shown that, at variance
with these predictions, solar-type stars do deplete Li on the MS.
Depletion  starts at $\sim 100$~Myr and smoothly goes on up to
the Hyades age, on a timescale of $\sim 1.4$~Gyr. After that age depletion
becomes bimodal: it is very fast for a fraction of stars, like the Sun itself,
and the lower envelope of M~67 (Pasquini et al. \cite{pas97} and references
therein), while 
depletion completely stops after
$\sim 1$~Gyr for a significant fraction of stars (Sestito \& Randich
\cite{sr05}; Prisinzano \& Randich~\cite{pr07}) 
In particular, Randich et al.
(\cite{randich_188} -RSP03
hereafter) have shown that members of the old (6--8~Gyr) OC
NGC~188 have a Li abundance only  a factor of two smaller than their
counterparts in  the factor of 10 younger Hyades; moreover, the average
value of Li in NGC~188
is surprisingly close to what is measured in Pop.~{\sc ii} stars.
This open cluster has a solar metallicity and, as discussed by RSP03, this
result might be a mere coincidence; however, the issue is obviously
worth  further  studies. It might indeed represent a link between
Pop.~{\sc i} and Pop.~{\sc ii}. Li depletion histories and provide a
clue to understand  whether the latter  have undergone Li depletion.
Investigating Li in additional very old OCs is thus very important.

It is worth mentioning that, to explain the unexpected Li
depletion during the MS,
several non-standard processes
were included in the models. The proposed mechanisms include
diffusion (Michaud~\cite{mich86}, Michaud et al.~\cite{mich04}; 
Chaboyer et al.~\cite{chab95}),
meridional circulation (Charbonnel \& Talon~\cite{ct99} and references
therein), angular momentum
loss and rotationally driven mixing (Eddington 1925;
Zahn 1974, 1992; Deliyannis \& Pinsonneault 1997),
gravity waves (Garc\'\i a L\'opez \& Spruit 1991; Montalb\`an \& Schatzmann
2000),
tachocline (Spergel \& Zahn 1992; Brun et al. 1999; Piau et al. 2003),
and combinations of waves and rotation (Charbonnel \& Talon~\cite{ct05};
Talon~\cite{talon}).
Each of these models makes specific predictions on the timescales of Li
depletion which can be compared with observational patterns.

We present here Li observations in a very large sample of members of
Berkeley 32. This open cluster is one of the  oldest in the Galaxy (5~Gyr
--e.g. D' Orazi et al.~\cite{dorazi}) and its metallicity is a factor of 
two below
solar (Sestito et al.~\cite{sestito06}); therefore, it provides an 
ideal sample to further
investigate the issue of the convergence of Li at old ages. 

The paper is organized as follows. In Sect.~2 we describe the sample,
the observations, and data reduction. Data analysis and the results
are presented in Sects. 3 and 4. A discussion is provided in Sect.~5,
followed by conclusions in Sect.~6. 
\section{Sample and observations}
The observations were obtained with VLT/FLAMES. 
Specifically,  we used the fiber link to UVES
to acquire spectra of evolved stars (RGB and clump stars)
to be used for the determination of
the cluster chemical compositions (Sestito et al.~\cite{sestito06};
Bragaglia et al.~\cite{brag08}),
while the Giraffe spectrograph and Medusa fiber system were employed
to observe MS and/or turn-off (TO) cluster candidates as well as
a few subgiants, in order to derive
membership and information on Li.
\subsection{Targets}
Photometric surveys of Be 32 have been performed by Kaluzny \& Mazur
(\cite{km91}), Richtler \& Sagar (\cite{rs00}), Hasegawa
et al. (\cite{hase}), and D'~Orazi et al. (\cite{dorazi}). A re-analysis
of D'~Orazi et al. photometry was performed by Tosi et al.(\cite{tbc}).
All these studies agree on a cluster age of 5--6~Gyr, on a subsolar
metallicity, and on a reddening E(B--V) in the range 0.08 -- 0.16,
with a most likely value of 0.1--0.12
(see discussion in Tosi et al.). Based on UVES
spectra of nine cluster members Sestito et a.~(\cite{sestito06})
derived for the
cluster a metallicity [Fe/H]=$-0.29\pm 0.04$. So far no spectroscopic
studies of MS stars have been performed and membership is based only
on photometric criteria.
\begin{figure}
\psfig{figure=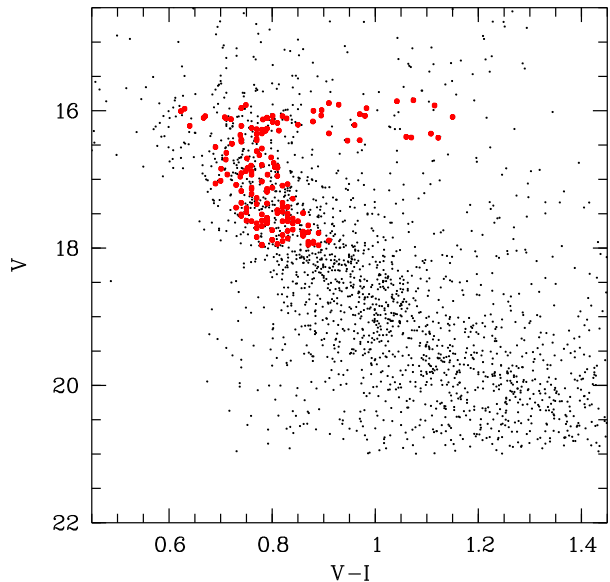, width=8.5cm}
\caption{V vs. V--I diagram of Be~32 with FLAMES targets shown
as filled symbols.
The whole catalog was constructed using Richtler \& Sagar (\cite{rs00})
photometry. We show in the figure only stars brighter than V=21. 
}\label{fig1}
\end{figure}
\begin{figure*}
\psfig{figure=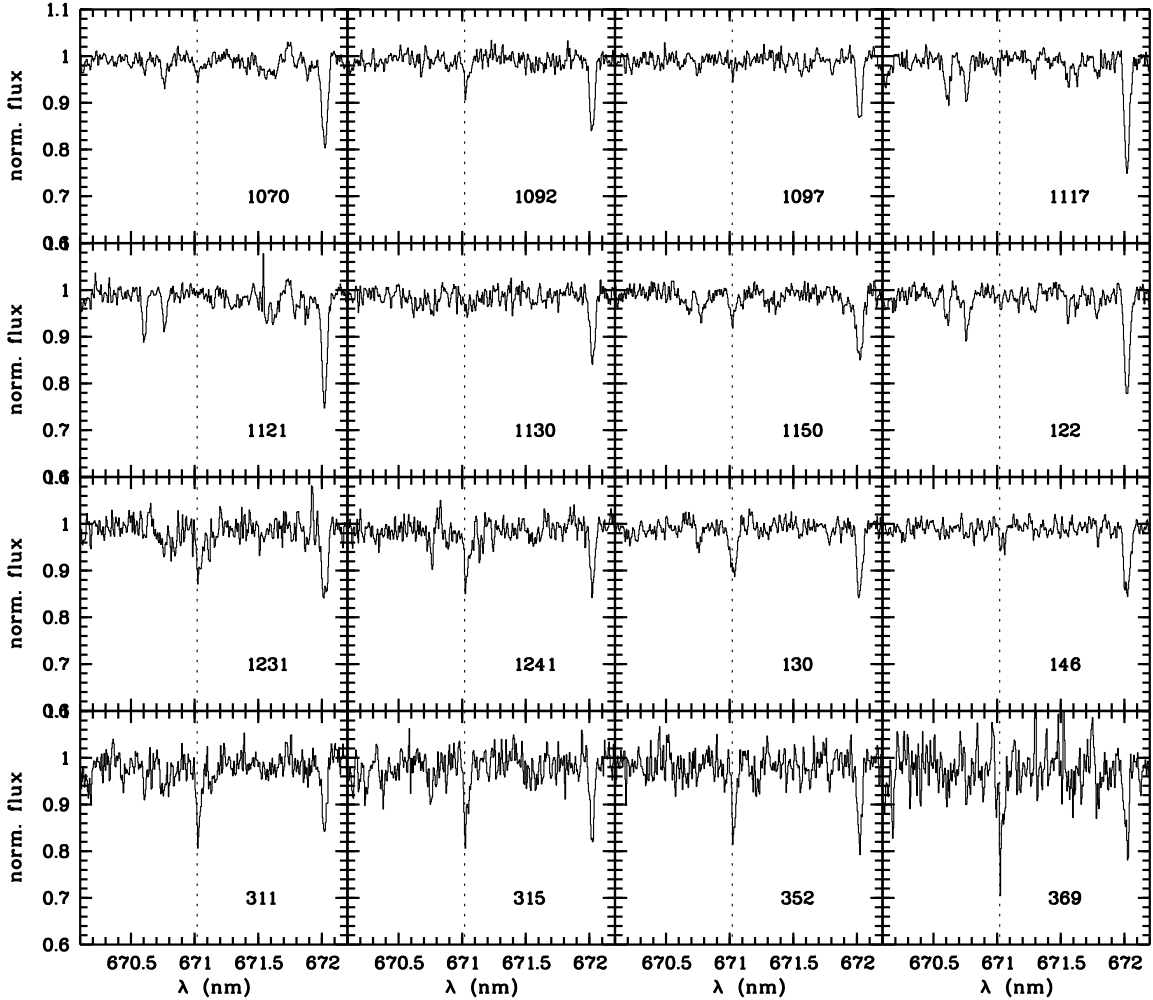, width=17cm}
\caption{Sample spectra in the Li region. The position of the Li line
is marked.}
\label{fig2}
\end{figure*}
We selected Giraffe target stars from the
catalogs of Kaluzny \& Mazur (\cite{km91}) and Richtler \& Sagar (\cite{rs00}),
since  the study of D'Orazi et al.
was still in preparation at the time of the observations.
The V -- V--I diagram of the target stars is displayed in Fig.~\ref{fig1}
which shows that the sample includes stars from the cluster subgiant branch and
turn-off (TO)
(V~$\sim 16$) down to V=18. 
As mentioned, membership for these candidates so far has been based 
only on photometry.
One of the purposes of this study is a more reliable
determination of membership via radial velocity measurements.
\subsection{Observations and data reduction}
Be~32 was observed with two different FLAMES configurations
(A and B) centered at 
RA(2000)=06h~58m~04.2s and DEC(2000)=$-$06d~28m~21.1s and
RA(2000)=06h~58m~02.0s and DEC(2000)=$-$06d~22m~41.4s, respectively.
We obtained four and one 3600~s long exposures for configurations A and B,
respectively.
Observations were obtained in Visitor mode 
on Jan. 20, 2005 (configuration A) and
Jan. 21, 2005 (configuration B).
Medusa fibers were allocated to 112 and 108 objects
in the two configurations, with 63 stars in common. In total we thus
obtained spectra of 157 cluster candidates. In both configurations 15 fibers
were put on the sky. Stars covered by configurations A and B 
were observed for a total of 4 and 1~hrs, respectively, while the
63 stars in common were observed for 5 hrs.
Target stars, together with their photometry, are listed in Table~\ref{sample}.
In the table we list our running ID (Col.~1), coordinates from Kaluzny \& Mazur
(\cite{km91}) if the star is present in their catalog otherwise from
Richtler \& Sagar (\cite{rs00} --Cols.~2-3); photometry from D'~Orazi
et al. (\cite{dorazi} --Cols.~4--7); 
V magnitude from Kaluzny \& Mazur
(\cite{km91}) if the star is present in their catalog otherwise from
Richtler \& Sagar (\cite{rs00} --Col.~8); B--V from Kaluzny \& Mazur
(Col.~9); V--I from Richtler \& Sagar (Col.~10); radial velocity
and membership flag (Cols.~11--12).
Giraffe was used
in conjunction with the 316 lines/mm grating and order-
sorting filter 15 (H15N) 
yielding a nominal resolving power R=17000 and covering a spectral
interval from 644.2 to 681.8~nm, including the 
Li~{\sc i}~670.8~nm line and several features to be used for radial velocity
measurements. 

Data reduction was performed using the Giraffe BLDRS
pipeline\footnote{version 1.12 -- http://girbldrs.sourceforge.net/},
following the standard procedure and steps (Blecha et al.~\cite{blecha04}). 
Sky subtraction was carried out separately; 
namely, we first computed the average sky (sky$_{\rm av.}$) 
of three sets of five sky spectra and then derived
the median of the three sky$_{\rm av.}$.
Radial velocities (RV) were
derived from each single spectrum (see below). Spectra of stars that were
not found to be RV~variables were then co-added.
Examples of final, sky-subtracted spectra are shown in Fig.~\ref{fig2}.
Final S/N range between 20 and 85, with an average value of $\sim 50$.
\section{Analysis}
\subsection{Radial velocities}
Radial velocities were obtained from our spectra and used 
to derive membership information for the sample stars. The 645 - 682 ~nm
region contains a large number of unblended lines (mainly Fe, Ca, Ti, and Ni)
of various strengths suitable for accurate RV measurements.
The individual lines used for the RV 
computation are listed in Table~\ref{lines}.
\begin{figure}
\psfig{figure=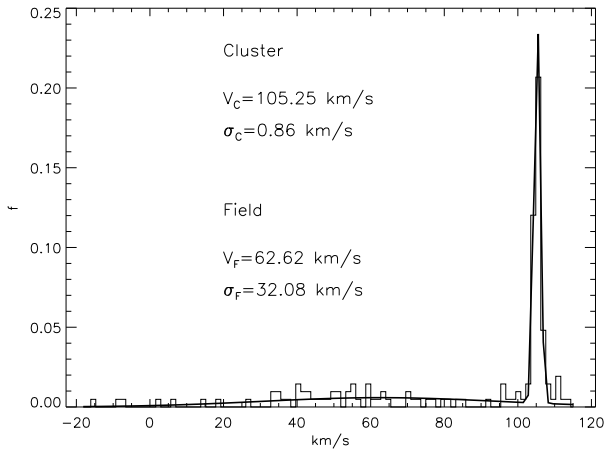, width=8.5cm}
\caption{Density distribution of radial velocities for the 153
stars for which we could get a measurement.
The two Gaussians (solid curves) indicate the best fits for the cluster and 
field, respectively. The average velocities along with $1 \sigma$ dispersion
are indicated.
}\label{fig3}
\end{figure}
\setcounter{table}{1}
\begin{table}
\caption{Individual lines suitable for RV computations.
\label{lines}}
\begin{tabular}{cc}
\hline
line \&  multiplet & $\lambda$ (\AA)\\
\hline
CaI 19	&	6449.810\\
CaI 18	&	6462.566\\
CaI 18	&	6471.660\\
FeI 206	&	6475.632\\
FeI 168	&	6494.985\\
FeI 268 + Ti 102 &6546.252\\
H$_\alpha$	&6562.808\\
FeI 268	&	6592.926\\
FeI 1197	&6633.764\\
NiI 43	&	6643.641\\
FeI 111	&	6663.446\\
FeI 268	&	6677.993\\
LiI 1	&	6707.815\\
CaI 32	&	6717.687\\
FeI 111	&	6750.152\\
NiI 57	&	6767.778\\
FeI 205	&	6783.710\\
FeI 268	&	6806.851\\
FeI 1197&	6810.280\\ \hline
\end{tabular}
\end{table}
The data analysis was performed by standard procedures within the IRAF
package\footnote{
IRAF is distributed by the National Optical
Astronomical Observatories, which  are operated by the Association of
Universities for Research in Astronomy,  under contract with the National
Science Foundation.}, 
fitting the strongest lines present in each spectrum by a Gaussian
profile. The resulting RVs from the individual lines were averaged, and
heliocentric corrections applied. Normally less than ten lines
per spectrum allowed us to obtain an RV value with an error of about 2 km/s;
RV values of spectra referring to the same star were then averaged for
stars in configuration A having multiple exposures.
Final radial velocities with their errors are given in Col.~9 of Table~
\ref{sample}. For stars with multiple RV measurements, the
error is the standard deviation from the average RV, while for stars in
configuration B the error is the uncertainty in the RV measurement itself.
\subsection{Li abundances}
Lithium abundances were derived by measuring the equivalent width (EW)
of the Li~{\sc I} line at 670.8~nm. Measurements were performed
by direct integration below the continuum.
For stars with detected Li, each measurement was performed
twice, meaning we determined the maximum and minimum  reasonable
values. We then
adopted the average between these two last values as EW measurement,
and we used half of the difference between
them as the error estimate on EW. In some RV members, the Li line could
not be detected and
we measured its upper limit, which was estimated
as the EW of the smallest detectable feature in the Li spectral region.
We measured the Li line in both radial velocity members and non-members.
Whereas for most radial velocity members we detected the Li line,
for six and 19 stars in fields A and B the S/N was
too low to even infer a meaningful upper limit. We mention in
passing that these stars are
not necessarily fainter than those where the Li line was measurable.
Lithium was also detected in 27 RV non-members.

Effective temperatures were determined on the basis of published
photometry (Richtler \& Sagar \cite{rs00}; D'Orazi et al. \cite
{dorazi}) and using the
color versus temperature calibrations by Alonso et al.~
(\cite{al96} --for MS stars, \cite{al99} --for evolved stars).
When available, we used the photometry of
D'Orazi et al., while we took colors from Richtler \& Sagar
for stars not included in the study of D'~Orazi et al.
As shown by Tosi et al.~(\cite{tbc}), the agreement between the two 
photometries is very good for most stars in the magnitude
range considered here. More specifically,
we used B--V colors
from D' Orazi et al. (\cite{dorazi}) for 45 stars out of the 47 for which
they were available.
Stars \#364 and \#1241 have B--V colors from D'Orazi et al., but
that of the former star 
is too red compared to the V--I, while B--V of the latter
is much bluer than the cluster sequence on the CM diagram; therefore
we did not use B--V from D'Orazi et al. We employed B--V of Kaluzny
\& Mazur (\cite{km91}) for two stars not included in the study
of D'Orazi et al. (\#236 and \#333) and for the aforementioned star \#364.
Finally, for star \#1241 and for eight stars
without available B--V we transformed 
V--I into B--V using a linear relationship that nicely
fits the sequence of stars with both colors available.
Reddening determinations towards Be~32 vary between
E(B--V)=0.10 and 0.18~mag (see Tosi et al.~\cite{tbc});
we assumed the value E(B--V)$=0.14$, determined
by Bragaglia et al.~(\cite{brag08}) using spectroscopic temperatures, since
we regard it as 
more reliable than values obtained from main sequence fitting.

To evaluate the random error that affects our temperature determinations, 
we compared
effective temperatures based on the B--V colors of D'~Orazi et al. with
those estimated from B--V of
Kaluzny \& Mazur for the  32 stars studied in both referenced papers.
The average
$\Delta$\teff~=(\teff$^{\rm D'Orazi}$-\teff$^{\rm K\&M}$)
is $-50$~K with a standard deviation
equal to 116~K. We adopt this value as the typical error
on effective temperatures. 

In previous studies (see Sestito\& Randich~\cite{sr05}), we derived
effective temperatures using the calibration of Soderblom et al.
(\cite{sod93a}). Since this calibration does not have a term
taking into account metallicity, we prefer to use Alonso's calibrations
for Be~32, whose metallicity is below solar.
Soderblom et al.'s calibration would have given warmer temperatures, with
a typical difference of 144~K. However, at solar metallicity
the two scales are very similar. 

As is well known,
at our resolution the Li line is blended with a Fe~{\sc i} line.
To estimate the contribution of Fe to the total EW,
we cannot use the analytical approximation of Soderblom et al.
(\cite{sod93b}), which was derived using stars with solar metallicity.
We instead estimated the EW of this feature employing 
MOOG (Sneden~\cite{sneden} 
--Version 2000) and the driver {\it ewfind} using the appropriate
metallicity and stellar parameters. Effective temperatures 
were derived as described above. The surface gravity for evolved stars
was estimated in the same fashion as in Sestito et al.~(\cite{sestito06}),
while for stars on the MS we assumed $\log$~g=4.5.
Microturbulence values $\xi=1.1$ and 1.5~km/s were used
for unevolved and evolved stars, respectively.
We found that for MS objects
EW(Fe~6707.44)=$(22.5-3.3\times 10^{-3}\times \rm T_{\rm eff}$)~\AA, while
the EW of the Fe line was computed separately for each evolved star.
The strength of this shallow Fe line has a very weak
dependence on both the microturbulence $\xi$ and $\log$~g: 
a change in $\xi$ of 1 km/s
results in a change in EW below 0.2~m\AA, while a change of 0.5~dex
in $\log$~g results in a difference of $\sim 0.2$~m\AA. 

The Li abundances were derived from corrected EWs using Soderblom 
et al. (\cite{sod93b}) curves of growth (COGs), which assume local
thermodynamic equilibrium (LTE). A correction for non--LTE effects following
Carlsson et al. (\cite{carls}) has
also been run on our data, but it produces no significant changes.
Li for evolved stars was instead derived using MOOG, which allows changing
surface gravity and microturbulence. Using of a different code and
method to derive Li for evolved and unevolved stars introduces
a small offset between the two abundance scales. By using MOOG, we would
obtain slightly higher \nli~for dwarf stars, typically by 0.05-0.1~dex.
On the one hand, this offset does not affect our results and,
in particular, our conclusions on dilution in more evolved stars
(Figs.~\ref{fig5} and \ref{fig6});
on the other hand, the use for Be~32 dwarf members of the same COGs
employed in
the literature for other OCs is critical to correctly draw
the time evolution of Li.

The final error on  the Li abundance measurement is  computed by
summing  quadratically the contributions from EW and temperature. 
Errors in the Fe I contribution due to errors in \teff~are
very small, much smaller than the errors in the EW measurements
themselves: namely, $\delta$EW(Fe~I)=0.5~m\AA~for $\delta$\teff=100~K.
\section{Results}
\subsection{Radial velocities and membership} \label{memb}
In Fig.~\ref{fig3}, we show the density distribution of 
radial velocities for
the 153 stars for which we were able to measure them. The figure
clearly shows a very narrow peak that indicates the presence
of the cluster. To derive the average cluster radial velocity,
we fitted the observed distribution
with two Gaussians, one for the cluster and
one for the field, and determined the best fit using a maximum
likelihood algorithm.
We obtained
RV$_{\rm cluster}=105.2$~Km/s, $\sigma_{\rm cluster}=0.86$ Km/s and
RV$_{\rm field}=62.6$~Km/s, $\sigma_{\rm field}=32.1$ Km/s, respectively.
Our measurement of the RV is in good agreement with the values of
D'~Orazi et al. (\cite{dorazi}, RV=106.7$\pm 8.5$~Km/s) from low-
resolution spectroscopy of 48 stars brighter than the turn-off,
of Sestito et al.~(\cite{sestito06}, RV=106$\pm 1.4$~Km/s), and
of Scott, Friel, \& Janes~(\cite{scott},
106$\pm 10$~Km/s) from the intermediate-resolution
spectroscopy of 10 giants. Also, three stars in our sample
have an RV measurement from D'Orazi et al.: \#74 (their \#698),
\#97 (their \#113), and \#1090 (their \#364). 
The agreement in RV is excellent for all of them. 
We stress that, thanks to our large
sample, we have been able to constrain the internal dispersion in
velocity much better than in previous studies, pushing it to below 1~Km/s.

We considered all stars as cluster members with RV within $3 \sigma$
from the cluster average: adopting this criterion, 59 and 24 stars
turned out to be members in fields A and B. 
The total fraction of members is
83/153 stars, i.e., 54 \%. 
The expected number of contaminants, or non members with RV consistent
with the cluster, is 2 stars. 
Since our observations were obtained
within the same night, we have only been able to identify one possible binary
systems (star \#1090). This system has average velocity consistent
with membership, but a double-line system and larger-than-average deviation
around the mean. We have tentatively
classified it as an SB2 cluster member.
Most likely, additional unidentified binaries
might be present among
stars with RV discrepant with membership and detected Li line. Further
follow-up is needed to confirm their membership. Therefore, we regard
the fraction of members as a lower limit to the real value.
\begin{figure*}
\includegraphics[height=17cm, angle=-90]{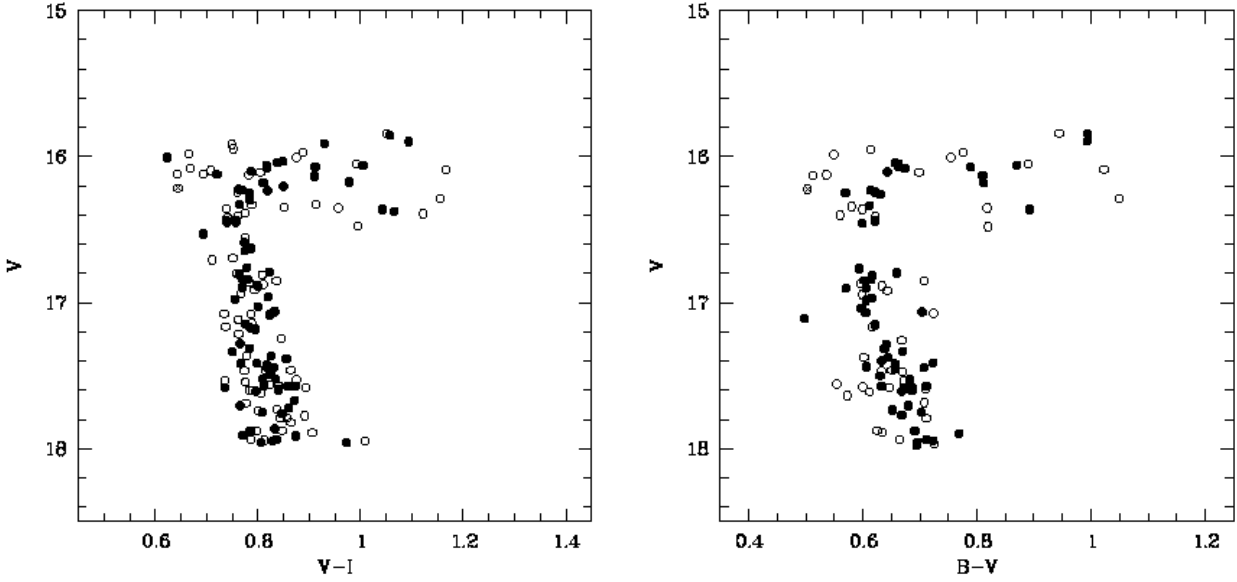}
\caption{``Cleaned" color-magnitude diagrams. 
Open symbols represent all observed
stars, while filled symbols denote radial velocity
members. The crossed symbol denotes the possible SB2 binary. When available
we have considered photometry from D'Orazi et al.~(\cite{dorazi}), while
in the other cases we have considered B--V colors from Kaluzny \& Mazur
(\cite{km91}) and V--I colors from Richtler \& Sagar (\cite{rs00}). 
}\label{fig4}
\end{figure*}
\begin{figure*}
\includegraphics[width=17cm,angle=0]{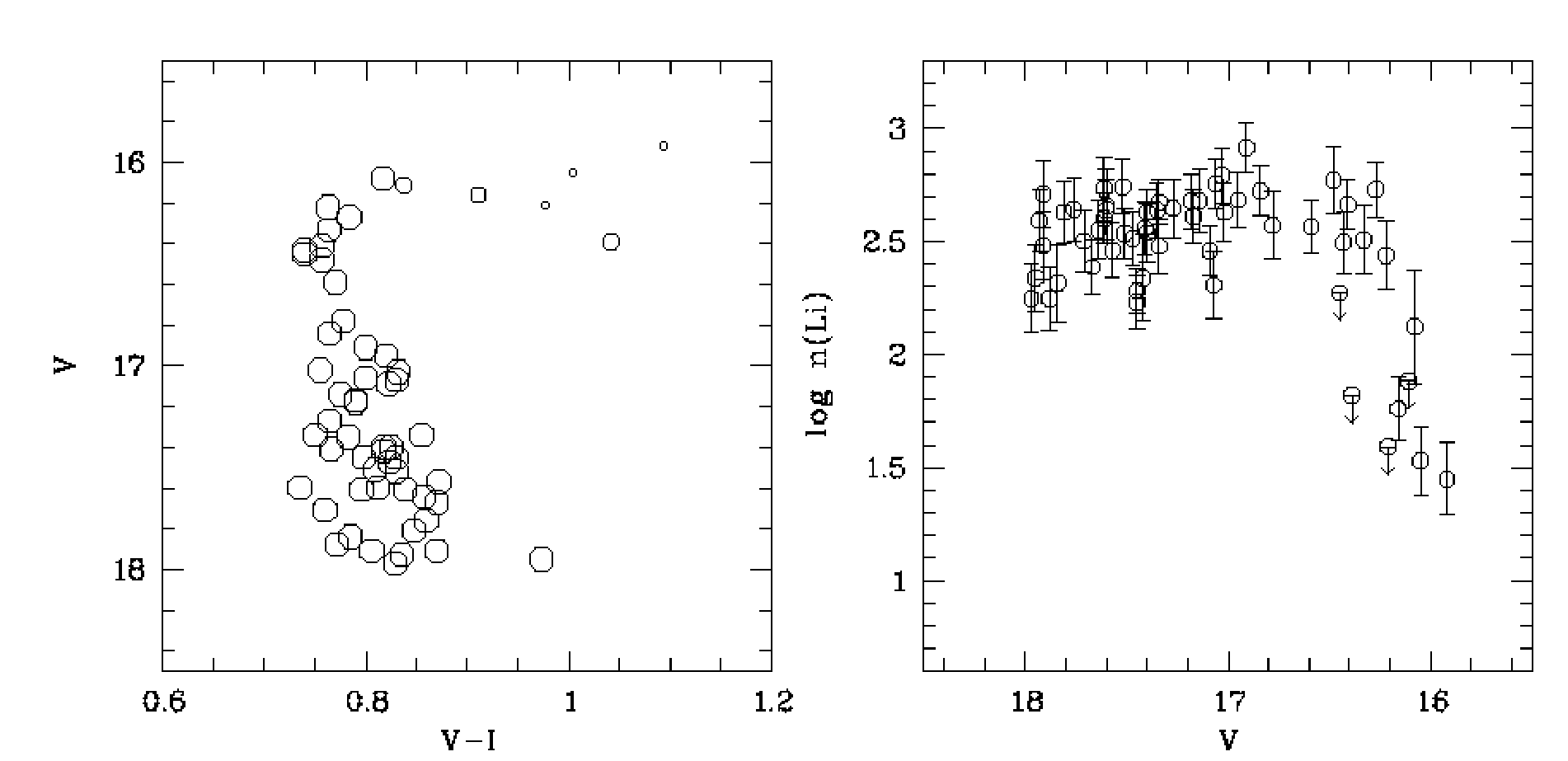}
\caption{
Panel a) V -- V--I diagram of confirmed cluster members. Different symbol
sizes indicate different Li abundance bins; namely, from the largest
to the smallest ones: \nli$\geq 2.0$, 
2.0$\leq$ \nli$< 1.7$, and \nli$<1.7$. For 47 of the 57 stars with
a Li measurement, we have considered photometry from D'Orazi et al. 
(\cite{dorazi}), while for the remaining 10 stars not included in the
D'Orazi et al. study, we used Richtler \& Sagar (\cite{rs00}) magnitudes
and colors.
Panel b): Lithium abundance --\nli, in the usual logarithmic scale
where n(H)=12--  as a function of V magnitude for confirmed cluster members.
}\label{fig5}
\end{figure*}
The ``cleaned" CM diagrams are shown in Fig. \ref{fig4}, where RV members
and non-members and the SB2 binary
are denoted with different symbols. Whereas
we refer to Tosi et al.~(\cite{tbc}) for a detailed
discussion of the analysis of the cluster CM diagram, we mention
here that our membership determination has allowed us to clean
the turn-off region, thus allowing a more solid derivation of cluster
parameters, as done by Tosi et al.
Among the subgiants we note the presence of two confirmed members 
(one in the V vs. B--V
diagram) somewhat fainter than the cluster sequence. We do not
have any explanation for them, but suggest that they might be the
two expected contaminants.
\subsection{Lithium abundances} \label{lithium}
In Table \ref{tab_lith} we list confirmed members (from Table~\ref{sample}) 
for which we were able to either measure the EW of the Li 
line or infer a reasonable upper limit. 
In the table we provide IDs (same as in Table
~\ref{sample}), dereddened B--V colors, effective temperatures, measured
Li equivalent widths, and Li abundances. The last are in the
usual notation \nli=$\log$N(Li)/N(H)+12.
In Fig.~\ref{fig5}a we
plot the V--\vmi~diagram of cluster members with different
symbol sizes denoting stars with different Li contents, while
in Fig~\ref{fig5}b we
show Li abundances as a function of V magnitude.
The figures very well illustrate the evolution of Li along
the CM diagram. All stars on the MS and at the TO 
have Li abundances larger than \nli$=2$ and
their present Li abundance is the result of MS depletion;
on the other hand, cluster members slightly more evolved than TO
have started diluting their surface Li, due to the deepening of
the convective zone (e.g., Randich et. al. \cite{randich_subg}).
The transition between stars that have and that have not undergone
dilution occurs in a very narrow magnitude range (see also Pasquini 
et al.~\cite{pas04}). Dilution 
progressively continues along the subgiant 
branch, up to a Li abundance $\sim 1$. 
\begin{figure}
\psfig{figure=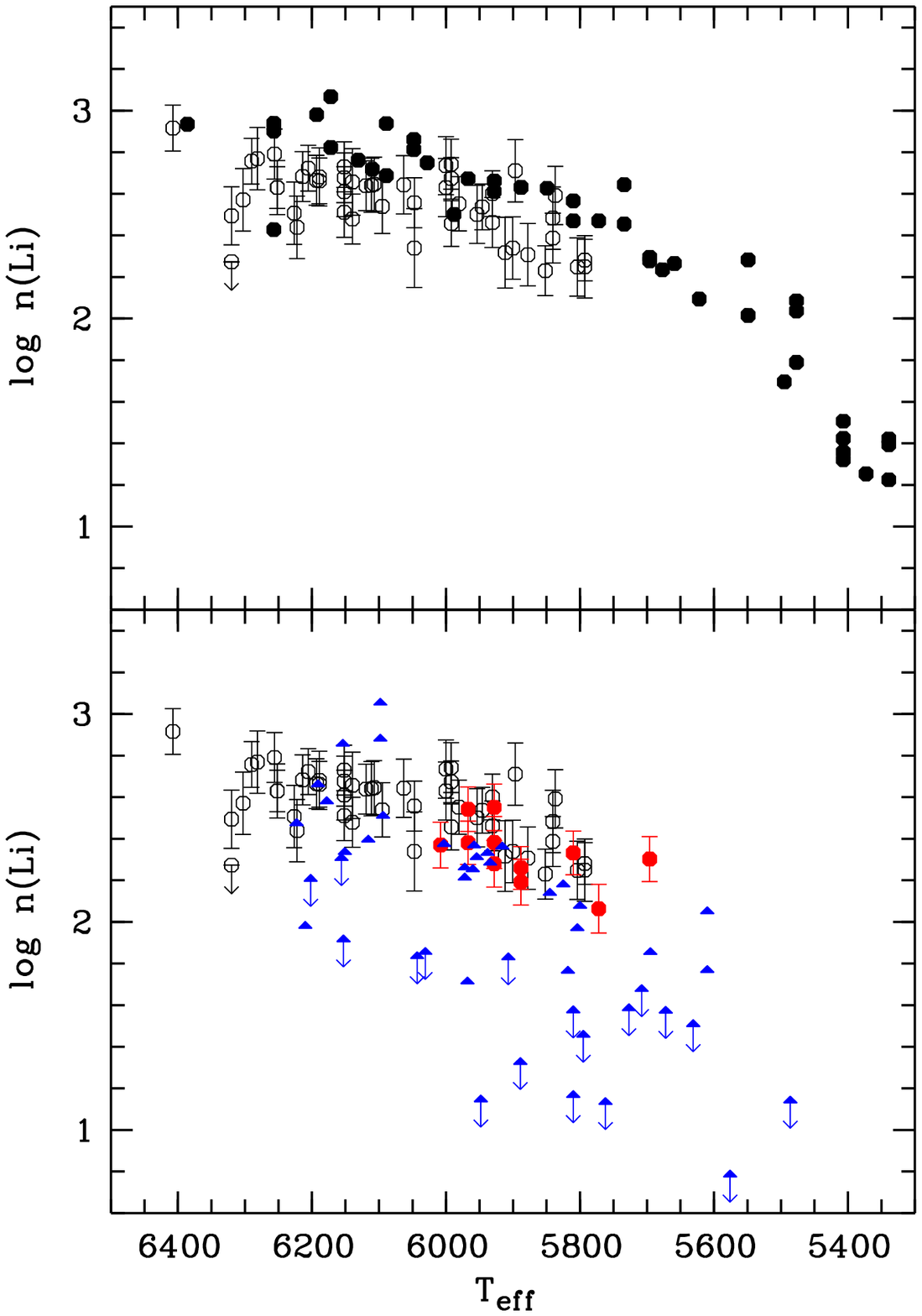, width=8.5cm}
\caption{Upper panel: comparison of the distribution
of \nli~as a function of \teff~for unevolved stars in Be~32 (open circles)
and the Hyades (filled circles); lower panel:
same as upper panel, but Be~32 is compared to
M67 (filled triangles) and NGC~188 (filled circles).
}\label{fig6}
\end{figure}
In Fig.~\ref{fig6} we show the usual plot of Li abundances
as a function of effective temperature for unevolved cluster stars (TO and
MS) and compare Be~32 with the much younger and more metal-rich Hyades
(upper panel) and with both the slightly younger M~67 and the 
slightly older cluster NGC~188 (lower panel). 
Both NGC~188 and M~67 OCs have a solar
metallicity (Randich et al.~\cite{randich_188},~\cite{randich_m67}), and thus
are a factor of $\sim 2$ more metal rich than Be~32. 
Lithium abundances for the three OCs were taken from the compilation
of Sestito \& Randich (\cite{sr05}). In that paper Li abundances
had been derived using the same COGs and NLTE correction employed here.

The figure shows that, at variance with the Hyades and M~67, but
more like NGC~188, Be~32 distribution does not show any major
trend of decreasing Li abundance with decreasing \teff.
We find average Li abundance
values of \nli$=2.63\pm 0.17, 2.62 \pm 0.07$, and $2.47 \pm 0.16$ 
for stars in the temperature ranges \teff~$>6200$ K, 6200$\geq$\teff$\geq 6000$
~K, and \teff$<$6000~K. The three average values are, within the errors, 
the same. Also,
at variance with M~67, but similar to NGC~188,
the Be~32 sample is not characterized by
any significant (larger than errors) dispersion in Li. 
More specifically, Figs.~\ref{fig5}b and \ref{fig6} show that the maximum
star-to-star difference for unevolved stars is on the order of 0.5~dex. 
Assuming Gaussian statistics,
this corresponds to a 1$\sigma$ dispersion of $\sim 0.15$~dex, 
comparable to the average error in \nli.
Given the large size of our sample, we regard the lack of a significant
scatter as real and
not due to low number statistics and conclude that the appearance
of the spread among open cluster stars is an exception rather than the rule,
since the majority of OCs do not show it.
Most important, whereas Hyades stars warmer than $\sim 6000$~K are somewhat
more Li-rich than their Be~32 counterparts, the Li distributions of cooler
stars are almost indistinguishable. The \nli--\teff~of NGC~188 
and Be~32
patterns are also very similar and close to the
upper envelope of M67. In other words, the comparison 
of OCs of different ages and metallicity
shows that, with the exceptions of the severely Li-depleted stars in the
lower envelope of M~67, stars in the $\sim 6000-5700$~K \teff~range seem to
converge to a very similar value of Li abundance.
\section{Discussion}
\subsection{Pop.~{\sc i} plateau}
\begin{figure}
\psfig{figure=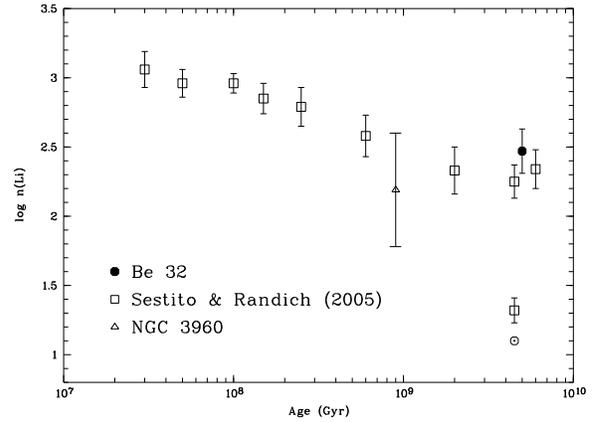, height=8.5cm, angle=-90}
\caption{
Average Li abundance as a function of age for stars in the
5750--6050~K~interval. Open squares are OCs
from Sestito and Randich (\cite{sr05}), the open triangle
represents the average for NGC~3960 (Prisinzano
\& Randich (\cite{pr07}), while the filled circle denotes Be~32.
For M~67 
the average of the upper and lower
envelopes are plotted. The Sun is also shown in the figure.
}\label{fig7}
\end{figure}

The present study allows us to make
firm conclusions about the empirical evolution of Li abundance during the MS
of stars with temperatures (but not necessarily masses --see next
section) similar to the Sun. 
In Fig.~\ref{fig7} we show the mean Li abundance as a function
of age for stars in the 6050--5750 temperature range. The figure
was done using the data of Sestito \& Randich (\cite{sr05}), to
which we added the average Li abundance of the $\sim$1~Gyr old
NGC~3960 (from Prisinzano
\& Randich \cite{pr07}) and the average for Be~32 inferred here. 
As already discussed by Sestito \& Randich, stars in this \teff~interval
undergo a smooth, but continuous Li depletion 
during the first $\sim 600$~Myr on the main sequence,
on a typical timescale
of $\sim 1.4$~Gyr. As already mentioned, at 
older ages depletion becomes bimodal: it continues for a fraction of stars,
while it stops for the majority
of stars and the average \nli~converge to a plateau, 
which is quantitatively -and surprisingly- close to the plateau
of Pop.~{\sc ii} stars. The evidence for this Pop.~{\sc i} 
plateau is statistically
confirmed by the inclusion in the sample of the Be 32, for which
we derived an average abundance \nli=$2.47 \pm 0.16$. This value is, slightly
higher, but within
the margins of error consistent, with that of
the intermediate age OCs ($2.33 \pm 0.17$),
the upper envelope of M~67 ($2.25 \pm 0.12$), and NGC~188 ($2.34 \pm 0.14$).
As to the Pop.~{\sc ii} plateau, values range between a minimum of
$2.10 \pm 0.09$ (Bonifacio et al.~\cite{bon07}) and a maximum of $\sim 2.4$
(Mel\'endez \& Ramirez~\cite{mel04}).
The conclusion is that
Pop.~I stars are not necessarily heavily Li-depleted, even at very old ages. 
The major consequence is that
Li cannot be used as an age indicator for stars older
than $\sim 500$~Myr: a lithium abundance in the interval
\nli~$\sim 2.3-2.6$ does not allow discerning whether
a star is 0.5 or 6~Gyr old. 
The above points are to be kept in mind when deriving the properties, age
in particular, of stars hosting extra-solar planets.
On the other hand, the Sun represents an exception and is
not representative of Li depletion in solar--type stars, since
it has undergone a larger-than-normal depletion.
We believe that low Li (a factor greater than 30-50 depletion) should
be regarded as indicative of a peculiar, but similar-to-the Sun, evolution.
\subsection{Lithium as a function of stellar mass}
Figure~\ref{fig6} clearly shows that the amount of Li depletion at a given
\teff~does not depend
on the cluster metallicity. The lack of any Li-metallicity dependence
has already been discussed by Sestito \& Randich (\cite{sr05}) on empirical
grounds and
by Piau et al.~(\cite{piau}) on theoretical grounds.
The latter study show that metallicity variations on
the order of 0.1~dex result
in changes of both \teff~and temperature at the basis
of the convective zone (T$_{\rm BCZ}$); however, the relationship between
\teff~and T$_{\rm BCZ}$, and thus the amount of depletion at a given \teff,
remain almost unaltered. The inclusion of
Be~32 in the open cluster sample and the comparison with the Hyades shows
that this holds true for metallicity differences as large
as 0.4-0.5~dex. 

As is well known, however, metallicity does affect stellar structure: 
specifically, for a given mass, lower 
metallicities correspond to higher \teff.
In contrast, for a given effective temperature, lower metallicity
stars have lower masses and stars with the same, close-to solar
temperature, do not have the same, solar mass, if their metal content
is different. In order to investigate Li depletion as a function
of mass, we derived masses for unevolved stars in Be~32, as well as for the 
Hyades, M~67, and NGC~188 using the Padova isochrones
(Girardi et al.~\cite{girardi} --http://pleiadi.pd.astro.it/) and
the appropriate metallicities. When needed, we interpolated between different
isochrones. 
\begin{figure}
\psfig{figure=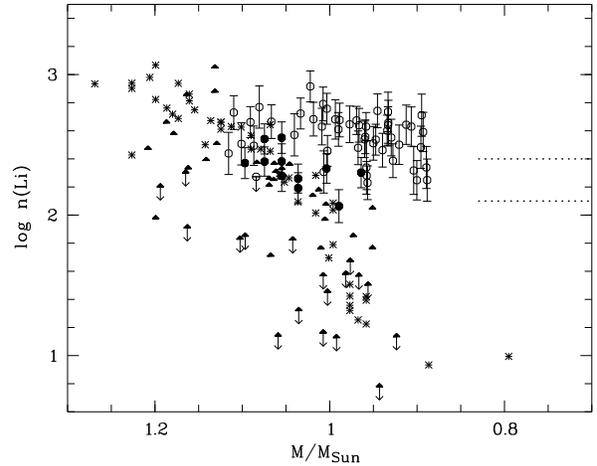, height=8.5cm, angle=-90}
\caption{
\nli~as a function of mass for Be~32 (open circles), 
the Hyades (asterisks),
M~67 (filled triangles), and NGC~188 (filled circles).
The horizontal lines delimits the range covered by Pop.~II stars considering
the lowest and highest values of the plateau.
}\label{fig8}
\end{figure}
In Fig.~\ref{fig8} we compare Li abundances as a function of
mass for unevolved stars
in Be~32 with the Hyades, M~67, and NGC~188 members. 
The range covered by Pop.~II stars is also shown.
This figure indicates that the distributions of the different OCs 
no longer overlap in the \nli-- mass
diagram as it was instead the case
for the \nli--\teff~plane. This is due to the fact that
Be~32 stars with temperatures close to solar are less 
massive than the Sun, of their counterparts in the solar metallicity
NGC~188 and M~67, and of their metal-rich counterparts in the Hyades.
The figure proves that Be~32 members at all
masses have depleted less Li than stars with similar mass
in the other old OCs; also, focusing on star with mass close to solar,
the figure shows that NGC~188 members and the upper envelope of M~67
have depleted the same Li as their more metal-rich Hyades counterparts.
In summary, both the
timescales and the amount of Li depletion are 
different for stars with the same mass and different metallicities.
This result indicates that, when looking at masses, 
metallicity affects the amount of depletion at a given age. In contrast,
we stress again that
the evolution of star with the same \teff~does not depend on [Fe/H]
(at least within $\pm 0.2-0.3$~dex from the solar value), since more
metal-poor and less massive stars, have the same internal structure of
solar-metallicity, solar-mass ones. 
\subsection{Comparison with model predictions}
\begin{figure}
\psfig{figure=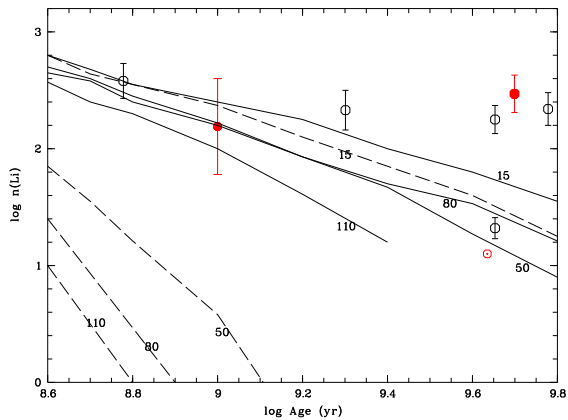, height=8.5cm, angle=-90}
\caption{
Same as Fig.~\ref{fig7}, but a narrower age interval is shown.
The empirical distribution is compared with the predictions
of the models by Charbonnel \& Talon~(\cite{ct05}). The latter have been
reconstructed starting from Fig.~2 in that paper. Different curves are
for different initial rotational velocities (in km/s) as labeled. Solid curves
correspond to model including waves + rotation, while dashed lines
denote models including only waves.
}\label{fig9}
\end{figure}

For a formally correct 
comparison of the empirical evolution of Li with model predictions at a
given mass, one should not mix data of OCs with significantly different
metallicities or, alternatively, the appropriate metallicity should
be considered. 
However, we believe that the comparison between model predictions
and empirical evolution in a given temperature range can still be performed,
since, at least as far as Li is concerned, the evolution of stars with
similar temperatures, but different masses and metallicities is virtually
the same. As an example,
we show again in Fig.~\ref{fig9} the
mean Li abundance as a function of age (see Fig.~\ref{fig7}) and compare
it with the predictions
of the models by Charbonnel \& Talon (\cite{ct05}), which include both 
rotational mixing and gravity waves. The figure indicates that these models
reproduce the observed distribution up
to 1~Gyr rather well. Also, the models with initial rotational velocity 
50--80~km/s
are in good agreement with the datapoints corresponding to the lower envelopes
of M~67 and with the solar datapoint. However,
even the model with the lowest initial rotation is not able to
fit the plateau in Li. In conclusion, while all classes of models including
extra-mixing processes predict that, once they start causing Li depletion,
they continue being efficient throughout the permanence on the MS, 
observational evidence indicates that this is not the case at all.
To our knowledge, none of the models proposed so far
predicts the convergence of Li at old ages.
We note that, when comparing model predictions with observations,
we assumed that all OCs have the same initial (meteoritic)
Li abundance. The assumption that the initial Li is instead
higher in young OCs 
would imply a different normalization of theory vs. models and a better
fit of the very old OCs; still, the disagreement between the observed
convergence and the models, which keep decreasing at old ages, remains.
\subsection{Pop.~I and Pop.~II plateaus}
In the previous sections we have definitively shown that stars
with temperatures similar to the Sun are not necessarily heavily Li-depleted
like the Sun is; instead, Li abundances of the majority of stars converge
towards a plateau, whose value, on the order of \nli$=2.3-2.5$, is close
(although not identical)
to that of Pop.~II stars. However, the path/masses/timescales are
different towards virtually this same average value of Li.
The similarity of the plateau of Pop.~I stars to the plateau of the
significantly more metal-poor Pop.~{\sc ii} is indeed very intriguing
and should be investigated on theoretical grounds.
Figure~\ref{fig8} indeed suggests that there might be a sequence
between old, solar-metallicity OCs and metal--poor less
massive Pop. II stars, with Be~32 in the middle. 
This sequence is purely phenomenological and it does not provide insight
into the depletion history of Pop.~II stars or into whether
they actually depleted some Li or not; nevertheless,
it is tempting to interpret it
as evidence that, whatever the initial Li abundance and whatever
the mixing mechanism, the final Li abundance is the same for metal-poor
Pop~{\sc ii} stars and more metal-rich ones. 
Those stars, like the Sun,
that instead deplete a much larger amount of Li represent an exception.
\section{Conclusions}
We carried out a FLAMES/Giraffe survey of almost 160 candidate
members of the old, metal-poor open cluster Be~32, with the goals of 
inferring membership 
and of studying the Li abundance pattern. To this aim,
we derived radial velocities
and Li abundances. Slightly more than half of the sample stars are confirmed
as cluster members. This is a lower limit, since we may have missed some 
binaries due to the too short time coverage of our observations. 
The Li versus \teff~distribution of unevolved members
overlaps with that of the slightly older, more metal-rich NGC~188,
and with the upper envelope of M~67. At variance with the last, Be~32 does
not show any dispersion in Li. The average abundance of stars with solar--like
{\it temperature} 
is slightly below the Hyades but, within the margins of error, 
the same as for their counterparts in the intermediate 
age OCs, the upper envelope of M~67, and NGC~188. 
This confirms, on solid and statistically significant grounds, that, 
with exception of Li--poor stars like the Sun, Li abundances
in solar-like Pop.~I stars converge towards a plateau value, implying 
that Li cannot be used as an age indicator after about 0.5~Gyr. In addition, 
the plateau of old OCs is close to the plateau
value of Pop.~II stars.
To our knowledge, none of the models including extra-mixing 
developed so far predicts the convergence toward a plateau at old ages.

Given their lower metallicity, Be~32 stars with temperature close to the Sun
are less massive than the Sun. A comparison in a \nli--mass diagram
shows that these stars have depleted much less Li than the Hyades with similar
mass. 
Our conclusion is that stars with different masses
and metallicities, but very similar temperatures, converge at old ages to
the same Li abundance, though following different Li depletion histories.
This might also be true for halo stars.
\begin{acknowledgements}
We are grateful to Paolo Span\`o
for help with the maximum-likelihood analysis of radial velocities
and to Paolo Montegriffo
for providing the software to cross-correlate catalogs. We thank the referee,
Andreas Korn, for very useful suggestions.
This work has made extensive use of the
services of WEBDA, ADS, CDS, etc. 
S. Randich has been supported by an INAF grant on 
{\it Young clusters as probes of star formation and early stellar evolution.}
\end{acknowledgements}
{}
\small
\setcounter{table}{0}
\begin{longtable}{rccccccccccrc}
\caption{Sample stars.
} \label{sample}\\
\hline
 ID & RA & DEC. & Conf. & B$_{\rm DBT}$ & V(B--V)$_{\rm DBT}$ & V(V--I)$_{\rm DBT}$ & I$_{\rm DBT}$ & V$_{\rm lit.}$ & B--V$_{\rm KM}$ & V--I$_{\rm RS}$ & v$_{\rm rad}$ & mf\\
  &\multicolumn{2}{c}{J2000}    &   &  &  &  &  & &  &  &  (Km/s) &   \\
\hline
\endfirsthead
\caption{continued.}\\
\hline
 ID & RA & DEC. & Conf. & B$_{\rm DBT}$ & V(B--V)$_{\rm DBT}$ & V(V--I)$_{\rm DBT}$ & I$_{\rm DBT}$ & V$_{\rm lit.}$ & B--V$_{\rm KM}$ & V--I$_{\rm RS}$ & v$_{\rm rad}$ & mf\\
  &\multicolumn{2}{c}{J2000}    &   &  &  &  &  & &  &  &  (Km/s) &   \\
\hline
\endhead
\hline\endfoot
      74 	 & 6 58 05.410 & 6 25 40.18 & B	&   16.840 &   15.846 &   15.855 &   14.798 &   15.862 &    0.956 &    1.042 &  105.4 $\pm 1.5$ & m \\
      77 	 & 6 57 56.856 & 6 24 26.47& B 	&   --- &   --- &   --- &   --- &   15.889 &    0.871 &    0.910 &  --- & ? \\
      84 	 & 6 58 09.005 & 6 28 55.83 & A	&   16.567 &   15.953 &   15.948 &   15.196 &   15.955 &    0.598 &    0.742 &   $44.7\pm 2.4$ & n \\
      86 	 & 6 58 05.060 & 6 24 59.35 & A	&   16.536 &   15.987 &   15.983 &   15.317 &   15.972 &    0.544 &    0.628 &  $111.5\pm 0.94$ & n \\
      91 	 & 6 58 00.353 & 6 27 44.37 & A	&   16.745 &   15.969 &   15.970 &   15.082 &   16.001 &    0.744 &    0.883 & $33.6\pm 0.14 $& n \\
      97 	 & 6 58 14.083 & 6 26 43.18 & A	&   16.930 &   16.060 &   16.060 &   15.056 &   16.052 &    0.872 &    0.975 &  $104.8\pm 0.43$ & m \\
      99 	 & 6 58 00.174 & 6 26 39.60 & B	&   16.862 &   16.073 &   16.073 &   15.161 &   16.069 &    0.781 &    0.895 &  $107.1\pm 1.2$ & m \\
     101 	 & 6 58 06.636 & 6 24 12.65 & B	&   16.938 &   16.048 &   16.051 &   15.059 &   16.074 &    0.867 &    0.980 &  $34.1\pm 0.9$ & n \\
     106 	 & 6 58 04.028 & 6 24 08.09 & A	&   16.807 &   16.108 &   16.108 &   15.303 &   16.099 &    0.717 &    0.786 &  $50.1\pm 0.43$ & n \\
     109 	 & 6 58 18.702 & 6 27 03.72 & A	&   16.711 &   16.049 &   16.041 &   15.203 &   16.106 &    0.668 &    0.794 &  $104.2\pm 0.32$ & m \\
     110 	 & 6 57 54.792 & 6 28 22.07 & B	&   16.698 &   16.042 &   16.032 &   15.183 &   16.112 &    0.649 &    0.828 &  $106.4\pm 2.2$ & m \\
     111 	 & 6 57 53.981 & 6 24 23.93 & B	&   16.749 &   16.106 &   16.102 &   15.315 &   16.120 &    0.620 &    0.805 &  $106.3\pm 2$ & m \\
     117 	 & 6 58 13.277 & 6 26 02.88 & B	&   16.943 &   16.133 &   16.136 &   15.225 &   16.156 &    0.766 &    0.879 &  $104.7\pm 1.2$ & m \\
     122 	 & 6 58 14.152 & 6 29 31.54 & A	&   16.992 &   16.180 &   16.176 &   15.199 &   16.207 &    0.823 &    0.955 &  $105.3\pm 0.25$ & m \\
     127 	 & 6 58 08.191 & 6 24 35.48 & A	&   16.868 &   16.247 &   16.243 &   15.483 &   16.249 &    0.603 &    0.755 &  $96.5\pm 1.25$ & n \\
     130 	 & 6 57 56.913 & 6 27 16.87 & A	&   16.890 &   16.260 &   16.254 &   15.470 &   16.267 &    0.618 &    0.765 &  $105.0\pm 0.28$ & m \\
     137 	 & 6 58 15.058 & 6 26 40.60 & B	&   17.335 &   16.286 &   16.291 &   15.136 &   16.333 &    1.019 &    1.108 &  $49.7\pm 1.4$ & n \\
     146 	 & 6 58 03.057 & 6 24 31.69 & A	&   17.062 &   16.441 &   16.437 &   15.680 &   16.414 &    0.629 &    0.766 &  $105.2\pm 0.32$ & m \\
     147 	 & 6 58 17.988 & 6 29 43.94 & A	&   16.961 &   16.401 &   16.388 &   15.612 &   16.426 &    0.586 &    0.744 &  $96.2\pm 0.59$ & n \\
     148 	 & 6 58 04.879 & 6 27 39.04 & A	&   17.033 &   16.411 &   16.406 &   15.646 &   16.435 &    0.588 &    0.770 &  $71.1\pm 0.16$ & n \\
     154 	 & 6 58 16.781 & 6 26 11.06 & B	&   17.058 &   16.459 &   16.452 &   15.695 &   16.481 &    0.583 &    0.722 &  $104.5\pm 2.4$ & m \\
     169 	 & 6 58 07.185 & 6 23 48.59 & B	&   17.206 &   16.626 &   16.620 &   15.891 &   16.613 &    0.579 &    0.711 &  --- & ? \\
     189 	 & 6 57 58.952 & 6 26 48.52 & B	&   17.458 &   16.798 &   16.792 &   15.970 &   16.762 &    0.662 &    0.803 &  $105.2\pm 1.4$ & m \\
     193 	 & 6 58 02.247 & 6 29 21.35 & A	&   17.362 &   16.768 &   16.759 &   15.981 &   16.785 &    0.605 &    0.780 &  $104.9\pm 0.84$ & m \\
     202 	 & 6 58 03.561 & 6 23 46.77 & B	&   17.459 &   16.845 &   16.838 &   16.058 &   16.835 &    0.617 &    0.752 &  $107.1\pm 3.1$ & m \\
     206 	 & 6 58 14.715 & 6 27 02.31 & B	&   17.431 &   16.814 &   16.808 &   16.044 &   16.844 &    0.604 &    0.701 &  $107.1\pm 2$ & m \\
     212 	 & 6 58 12.503 & 6 28 33.07 & B	&   17.469 &   16.873 &   16.863 &   16.079 &   16.879 &    0.605 &    0.749 &  $63.3\pm 2.1$  & n \\
     213 	 & 6 58 16.427 & 6 23 34.95 & B	&   17.450 &   16.848 &   16.841 &   16.072 &   16.881 &    0.605 &    0.760 &  $103.3\pm 1.9$ & m \\
     216 	 & 6 57 54.890 & 6 27 12.27 & A	&   17.473 &   16.903 &   16.888 &   16.088 &   16.909 &    0.570 &    0.741 &  $104.8\pm 1.3$ & m \\
     218 	 & 6 58 04.891 & 6 23 31.10 & B	&   17.543 &   16.944 &   16.936 &   16.169 &   16.929 &    0.627 &    0.713 &  $108.6\pm 3.6$ & n \\
     219 	 & 6 58 15.312 & 6 28 15.59 & A	&   17.520 &   16.886 &   16.877 &   16.067 &   16.929 &    0.632 &    0.760 &  $110.8\pm 0.8$ & n \\
     220 	 & 6 58 10.153 & 6 23 51.84 & B	&   17.561 &   16.918 &   16.913 &   16.120 &   16.930 &    0.633 &    0.791 &  $81.3\pm 3.0$ & n \\
     222 	 & 6 58 16.404 & 6 27 15.53 & B	&   17.509 &   16.903 &   16.895 &   16.125 &   16.960 &    0.571 &    0.738 &  $105.3\pm 2.4$ & m \\
     231 	 & 6 58 14.010 & 6 26 57.99 & A,B	&   17.589 &   16.983 &   16.977 &   16.222 &   17.019 &    0.589 &    0.695 &  $104.8\pm 1.0$ & m \\
     236 	 & 6 58 02.398 & 6 26 57.18 & A	&   --- &   --- &   --- &   --- &   17.065 &    0.704 &    0.833 &  $105.5\pm 0.57$ & m \\
     240 	 & 6 58 12.220 & 6 25 57.42 & A,B	&   17.797 &   17.073 &   17.078 &   16.292 &   17.124 &    0.658 &    0.804 &  $35.6\pm 1.1$ & n \\
     241 	 & 6 57 57.665 & 6 25 41.09 & A,B	&   17.773 &   17.152 &   17.146 &   16.371 &   17.135 &    0.591 &    0.759 &  $105.8\pm 0.54$ & m \\
     245 	 & 6 58 02.865 & 6 25 09.94 & A,B	&   --- &   --- &   --- &   --- &   17.167 &    0.616 &    0.737 &  $110.6\pm 0.4$ & n \\
     254 	 & 6 58 01.182 & 6 29 00.09 & A	&   17.925 &   17.257 &   17.249 &   16.403 &   17.279 &    0.674 &    0.835 &  $45.7\pm 1.0$ & n \\
     265 	 & 6 58 12.307 & 6 24 50.80 & A,B	&   18.004 &   17.334 &   17.336 &   16.586 &   17.342 &    0.679 &    0.737 &  $104.7\pm 1.2$ & m \\
     271 	 & 6 58 16.474 & 6 24 00.78 & A,B	&   17.955 &   17.317 &   17.312 &   16.529 &   17.355 &    0.641 &    0.770 &  $105.2\pm 1.11$ & m \\
     276 	 & 6 57 58.540 & 6 28 06.48 & A,B	&   18.020 &   17.376 &   17.368 &   16.543 &   17.396 &    0.649 &    0.823 &  $104.8\pm 0.32$ & m \\
     277 	 & 6 57 53.234 & 6 27 03.85 & A,B	&   18.046 &   17.440 &   17.427 &   16.609 &   17.398 &    0.513 &    0.831 &  $104.8\pm 0.7$ & m \\
     278 	 & 6 58 06.158 & 6 26 01.56 & A,B	&   18.075 &   17.419 &   17.418 &   16.652 &   17.410 &    0.643 &    0.734 &  $104.7\pm 1.92$ & m \\
     288 	 & 6 58 08.217 & 6 26 28.77 & A,B	&   18.137 &   17.414 &   17.417 &   16.619 &   17.447 &    0.636 &    0.806 &  $107.6\pm 0.6$ & m \\
     289 	 & 6 58 05.985 & 6 25 23.23 & A,B	&   18.154 &   17.447 &   17.445 &   16.615 &   17.454 &    0.670 &    0.807 &  $105.6\pm 0.9$ & m \\
     292 	 & 6 58 06.568 & 6 26 53.29 & A,B	&   18.103 &   17.470 &   17.465 &   16.691 &   17.485 &    0.579 &    0.752 &  $92.5\pm 1.0$ & n \\
     294 	 & 6 58 15.206 & 6 29 36.09 & A,B	&   18.115 &   17.464 &   17.458 &   16.644 &   17.491 &    0.677 &    0.814 &  $63.6\pm 1.0$ & n \\
     299 	 & 6 58 01.374 & 6 27 16.75 & A,B	&   18.206 &   17.524 &   17.521 &   16.712 &   17.512 &    0.645 &    0.777 &  $105.3\pm 1.3$ & m \\
     311 	 & 6 58 05.745 & 6 23 45.05 & A	&   18.263 &   17.577 &   17.569 &   16.697 &   17.573 &    0.686 &    0.828 &  $104.1\pm 0.5$ & m \\
     313 	 & 6 58 06.532 & 6 25 38.80 & A,B	&   18.282 &   17.571 &   17.580 &   16.845 &   17.597 &    0.635 &    0.751 &  $105.1\pm 1.1$ & m \\
     314 	 & 6 58 10.331 & 6 27 12.53 & A,B	&   18.208 &   17.575 &   17.566 &   16.754 &   17.602 &    0.616 &    0.794 &  $107.6\pm 0.6$ & m \\
     315 	 & 6 58 00.111 & 6 27 07.30 & A	&   18.287 &   17.601 &   17.596 &   16.757 &   17.607 &    0.600 &    0.822 &  $105.7\pm 1.3$ & m \\
     316 	 & 6 58 10.618 & 6 24 23.52 & A,B	&   18.276 &   17.608 &   17.606 &   16.810 &   17.610 &    0.670 &    0.761 &  $105.8\pm 0.5$ & m \\
     317 	 & 6 58 16.862 & 6 26 32.37 & A,B	&   18.301 &   17.591 &   17.584 &   16.691 &   17.612 &    0.694 &    0.850 &  $110.0\pm 1.6$ & n \\
     325 	 & 6 57 55.799 & 6 28 12.39 & A,B	&   18.257 &   17.584 &   17.575 &   16.717 &   17.635 &    0.671 &    0.840 &  $105.3\pm 0.6$ & m \\
     326 	 & 6 58 17.239 & 6 23 49.38 & A,B	&   18.222 &   17.610 &   17.601 &   16.812 &   17.645 &    0.625 &    0.771 &  $114.1\pm 1.13$ & n \\
     331 	 & 6 58 06.080 & 6 26 36.12 & A,B	&   18.392 &   17.684 &   17.688 &   16.911 &   17.679 &    0.692 &    0.784 &  $46.7\pm 2$ & n \\
     333 	 & 6 58 14.218 & 6 25 31.96 & A	&   --- &   --- &   --- &   --- &   17.707 &    0.680 &    0.765 &  $104.8\pm 0.7$ & m \\
     344 	 & 6 57 58.305 & 6 25 58.35 & A	&   18.503 &   17.792 &   17.790 &   16.946 &   17.768 &    0.707 &    0.873 &  $100.1\pm 1.0$ & n \\
     352 	 & 6 58 18.582 & 6 29 32.81 & A,B	&   18.436 &   17.768 &   17.760 &   16.913 &   17.808 &    0.678 &    0.819 &  $104.5\pm 1$ & m \\
     357 	 & 6 58 00.620 & 6 24 29.96 & A,B	&   18.569 &   17.878 &   17.879 &   17.094 &   17.839 &    0.677 &    0.773 &  $104.5\pm 1.5$ & m \\
     362 	 & 6 58 10.854 & 6 26 43.75 & A,B	&   --- &   --- &   --- &   --- &   17.878 &    0.625 &    0.797 &  $51.7\pm 5$ & n \\
     364 	 & 6 58 04.497 & 6 26 55.95 & A,B	&   18.666 &   17.897 &   17.909 &   17.138 &   17.880 &    0.723 &    0.803 &  $105.4\pm 1.2$ & m \\
     367 	 & 6 58 09.763 & 6 23 37.42 & A,B	&   18.523 &   17.889 &   17.876 &   17.029 &   17.896 &    0.658 &    0.820 &  $40.5\pm 0.3$ & n \\
     369 	 & 6 57 54.006 & 6 24 58.68 & A,B	&   18.650 &   17.955 &   17.955 &   17.149 &   17.910 &    0.755 &    0.879 &  $103.4\pm 1.0$ & m \\
     371 	 & 6 57 58.413 & 6 23 49.71 & A,B	&   18.652 &   17.941 &   17.939 &   17.103 &   17.934 &    0.723 &    0.877 &  $106.0\pm 0.8$ & m \\
     373 	 & 6 57 59.759 & 6 23 29.26 & A,B	&   18.602 &   17.937 &   17.935 &   17.149 &   17.952 &    0.609 &    0.777 &  $70.2\pm 1.5$ & n \\
     374 	 & 6 58 10.035 & 6 26 56.43 & A	&   18.670 &   17.947 &   17.947 &   17.118 &   17.965 &    0.669 &    0.775 &  $105.2\pm 0.7$ & m \\
     927 	 & 6 58 05.163 & 6 23 01.21 & A	&   14.268 &   12.889 &   12.904 &   11.461 &   12.896 &   --- &    1.425 &  $105.5\pm 0.3$ & m \\
    1043 	 & 6 58 11.169 & 6 22 14.52 & B	&   16.785 &   15.841 &   15.845 &   14.794 &   15.848 &   --- &    1.074 &  $37.2\pm 1.6$ & n \\
    1047 	 & 6 57 51.082 & 6 20 53.46 & B	&   --- &   --- &   --- &   --- &   15.912 &   --- &    0.929 &  $104.9\pm 2.3$ & m \\
    1049 	 & 6 58 03.201 & 6 21 12.02 & B	&   --- &   --- &   --- &   --- &   15.917 &   --- &    0.749 &  $43.5\pm 6.3$ & n \\
    1050 	 & 6 58 03.329 & 6 22 25.01 & B	&   16.887 &   15.894 &   15.899 &   14.805 &   15.923 &   --- &    1.115 &  $105.3\pm 1.2$ & m \\
    1051 	 & 6 58 28.829 & 6 20 59.65 & B	&   --- &   --- &   --- &   --- &   15.964 &   --- &    0.983 &  --- & ? \\
    1057 	 & 6 58 20.765 & 6 22 07.75 & B	&   16.761 &   16.007 &   16.006 &   15.131 &   15.991 &   --- &    0.896 &  $42.0\pm 2.1$ & n \\
    1059 	 & 6 57 44.117 & 6 24 57.58 & B	&   --- &   --- &   --- &   --- &   16.006 &   --- &    0.623 &  $106.8\pm 3.8$ & m \\
    1066 	 & 6 58 24.787 & 6 24 09.94 & B	&   16.756 &   16.082 &   16.077 &   15.260 &   16.077 &   --- &    0.801 &  $105.5\pm 1.2$ & m \\
    1069 	 & 6 57 52.859 & 6 29 07.52 & A	&   --- &   --- &   --- &   --- &   16.082 &   --- &    0.668 &  $99.5\pm 1.6$ & n \\
    1070 	 & 6 58 04.031 & 6 22 57.96 & A	&   16.728 &   16.066 &   16.060 &   15.243 &   16.083 &   --- &    0.818 &  $104.8\pm 0.3$ & m \\
    1072 	 & 6 58 23.376 & 6 27 00.93 & A	&   17.113 &   16.090 &   16.091 &   14.924 &   16.089 &   --- &    1.153 &  $51.8\pm 0.2$ & n \\
    1073 	 & 6 57 43.108 & 6 22 07.05 & B	&   --- &   --- &   --- &   --- &   16.095 &   --- &    0.708 &  $69.7\pm 5.6$ & n \\
    1074 	 & 6 58 19.457 & 6 22 01.50 & B	&   16.640 &   16.128 &   16.123 &   15.480 &   16.103 &   --- &    0.667 &  $58.9\pm 1.4$ & n \\
    1075 	 & 6 58 21.900 & 6 22 13.24 & A	&   16.662 &   16.126 &   16.119 &   15.425 &   16.108 &   --- &    0.711 &  $27.0\pm 1.1$ & n \\
    1077 	 & 6 58 26.626 & 6 30 12.54 & B	&   --- &   --- &   --- &   --- &   16.123 &   --- &    0.720 &  $106.5\pm 1.8$ & m \\
    1078 	 & 6 57 51.039 & 6 29 04.62 & B	&   --- &   --- &   --- &   --- &   16.127 &   --- &    0.782 &  $14.6\pm 5.2$ & n \\
    1082 	 & 6 58 27.554 & 6 24 16.90 & B	&   --- &   --- &   --- &   --- &   16.163 &   --- &    0.801 &  --- & ? \\
    1084 	 & 6 58 10.628 & 6 20 40.14 & B	&   --- &   --- &   --- &   --- &   16.177 &   --- &    0.810 &  $104.4\pm 1.7$ & m \\
    1088 	 & 6 57 52.147 & 6 29 18.22 & B	&   --- &   --- &   --- &   --- &   16.204 &   --- &    0.850 &  $105.8\pm 2.0$ & m \\
    1090 	 & 6 58 14.337 & 6 22 57.76 & A	&   16.726 &   16.223 &   16.217 &   15.573 &   16.216 &   --- &    0.638 &  $106.5\pm 4.5$ & m, SB2? \\
    1092 	 & 6 58 23.454 & 6 28 52.21 & A	&   16.844 &   16.231 &   16.225 &   15.462 &   16.220 &   --- &    0.744 &  $105.2\pm 0.6$ & m \\
    1097 	 & 6 58 10.641 & 6 30 41.64 & A	&   16.819 &   16.250 &   16.233 &   15.414 &   16.258 &   --- &    0.789 &  $104.6\pm 0.2$ & m \\
    1101 	 & 6 57 54.904 & 6 22 58.15 & B	&   16.858 &   16.238 &   16.232 &   15.462 &   16.275 &   --- &    0.780 &  $105.5\pm 1.3$ & m \\
    1102 	 & 6 58 15.248 & 6 21 37.54 & B	&   16.962 &   16.363 &   16.346 &   15.495 &   16.289 &   --- &    0.813 &  $84.7\pm 2.9$ & n \\
    1105 	 & 6 58 03.716 & 6 20 16.81 & B	&   --- &   --- &   --- &   --- &   16.300 &   --- &    0.784 &  $103.9\pm 1.7$ & m \\
    1112 	 & 6 58 20.650 & 6 31 00.58 & B	&   --- &   --- &   --- &   --- &   16.330 &   --- &    0.913 &  $39.9\pm 0.5$ & n \\
    1113 	 & 6 58 20.285 & 6 23 48.40 & A	&   16.947 &   16.335 &   16.329 &   15.565 &   16.333 &   --- &    0.775 &  $104.9\pm 0.7$ & m \\
    1114 	 & 6 57 51.759 & 6 27 17.88 & A	&   16.922 &   16.342 &   16.329 &   15.541 &   16.336 &   --- &    0.769 &  $101.7\pm 0.9$ & n \\
    1116 	 & 6 57 45.696 & 6 24 02.85 & B	&   --- &   --- &   --- &   --- &   16.354 &   --- &    0.738 &  $75.0\pm 5.9$ & n \\
    1117 	 & 6 57 51.055 & 6 31 13.59 & A	&   --- &   --- &   --- &   --- &   16.376 &   --- &    1.065 &  $105.0\pm 0.2$ & m \\
    1121 	 & 6 58 03.681 & 6 22 17.30 & A	&   17.255 &   16.362 &   16.361 &   15.319 &   16.386 &   --- &    1.067 &  $104.8\pm 0.4$ & m \\
    1122 	 & 6 58 24.188 & 6 21 04.34 & B	&   --- &   --- &   --- &   --- &   16.393 &   --- &    1.122 &  $43.4\pm 1.6$ & n \\
    1123 	 & 6 58 21.389 & 6 30 32.08 & A	&   --- &   --- &   --- &   --- &   16.420 &   --- &    0.741 &   $54.1\pm 0.4$ & n \\
    1126 	 & 6 57 59.971 & 6 30 05.96 & A	&   17.171 &   16.353 &   16.352 &   15.395 &   16.431 &   --- &    0.966 &  $17.9\pm 0.5$ & n \\
    1128 	 & 6 58 30.891 & 6 23 30.75 & B	&   --- &   --- &   --- &   --- &   16.434 &   --- &    0.738 &  $105.0\pm 1.0$ & m \\
    1129 	 & 6 57 50.796 & 6 27 03.39 & B	&   17.300 &   16.481 &   16.476 &   15.482 &   16.435 &   --- &    0.946 &  $58.7\pm 2.7$ & n \\
    1130 	 & 6 58 14.736 & 6 31 19.87 & A	&   --- &   --- &   --- &   --- &   16.450 &   --- &    0.740 &  $107.3\pm 0.4$ & m \\
    1141 	 & 6 58 20.329 & 6 30 36.64 & A	&   --- &   --- &   --- &   --- &   16.531 &   --- &    0.693 &  $105.2\pm 0.5$ & m \\
    1144 	 & 6 57 50.902 & 6 21 57.44 & A	&   --- &   --- &   --- &   --- &   16.555 &   --- &    0.776 &  $59.5\pm 0.3$ & n \\
    1150 	 & 6 58 10.245 & 6 21 06.25 & A	&   --- &   --- &   --- &   --- &   16.590 &   --- &    0.774 &  $106.1\pm 0.8$ & m \\
    1153 	 & 6 58 09.489 & 6 21 07.21 & B	&   --- &   --- &   --- &   --- &   16.629 &   --- &    0.787 &  $104.5\pm 2.9$ & m \\
    1156 	 & 6 57 51.474 & 6 22 38.53 & B	&   --- &   --- &   --- &   --- &   16.646 &   --- &    0.775 &  $106.0\pm 2.2$ & m \\
    1161 	 & 6 57 48.933 & 6 22 14.65 & B	&   --- &   --- &   --- &   --- &   16.695 &   --- &    0.751 &  $62.2\pm 3.0$ & n \\
    1164 	 & 6 58 25.577 & 6 29 39.85 & A	&   --- &   --- &   --- &   --- &   16.707 &   --- &    0.711 &  $95.9\pm 0.2$ & n \\
    1179 	 & 6 58 17.498 & 6 20 46.12 & B	&   --- &   --- &   --- &   --- &   16.799 &   --- &    0.759 &  $35.4\pm 4.4$ & n \\
    1181 	 & 6 57 54.593 & 6 21 30.34 & A	&   --- &   --- &   --- &   --- &   16.809 &   --- &    0.808 &  $-15.8\pm 1.0$ & n \\
    1187 	 & 6 58 21.059 & 6 24 31.50 & A	&   17.561 &   16.853 &   16.851 &   16.015 &   16.829 &   --- &    0.814 &  $34.9\pm 0.3$ & n \\
    1195 	 & 6 58 20.695 & 6 19 17.52 & B	&   --- &   --- &   --- &   --- &   16.880 &   --- &    0.798 &  $104.4\pm 1.7$ & m \\
    1212 	 & 6 58 06.229 & 6 29 45.40 & A	&   17.587 &   16.972 &   16.960 &   16.140 &   16.955 &   --- &    0.806 &  $104.2\pm 0.9$ & m \\
    1228 	 & 6 57 51.634 & 6 27 48.18 & A,B	&   17.675 &   17.070 &   17.056 &   16.224 &   17.028 &   --- &    0.776 &  $105.1\pm 0.8$ & m \\
    1231 	 & 6 57 57.805 & 6 30 26.93 & A,B	&   17.636 &   17.039 &   17.027 &   16.227 &   17.060 &   --- &    0.693 &  $105.8\pm 1.4$ & m \\
    1238 	 & 6 57 53.881 & 6 22 43.21 & A,B	&   --- &   --- &   --- &   --- &   17.077 &   --- &    0.734 &  $98.2\pm 1.5$ & n \\
    1241 	 & 6 57 52.746 & 6 26 48.95 & A,B	&   17.607 &   17.109 &   17.083 &   16.260 &   17.087 &   --- &    0.816 &  $105.4\pm 0.6$ & m \\
    1245 	 & 6 58 08.660 & 6 21 11.52 & A,B	&   --- &   --- &   --- &   --- &   17.116 &   --- &    0.762 &  $40.3\pm 0.9$ & n \\
    1250 	 & 6 57 46.565 & 6 23 34.51 & A,B	&   --- &   --- &   --- &   --- &   17.143 &   --- &    0.789 &  $64.7\pm 0.8$ & n \\
    1257 	 & 6 58 06.810 & 6 21 30.85 & A,B	&   --- &   --- &   --- &   --- &   17.172 &   --- &    0.785 &  $105.1\pm 0.3$ & m \\
    1263 	 & 6 58 11.569 & 6 31 06.92 & A,B	&   --- &   --- &   --- &   --- &   17.183 &   --- &    0.794 &  $104.4\pm 1.5$ & m \\
    1265 	 & 6 58 25.849 & 6 20 00.02 & A,B	&   --- &   --- &   --- &   --- &   17.214 &   --- &    0.763 &  $48.9\pm 1.2$ & n \\
    1279 	 & 6 58 19.836 & 6 24 48.43 & A,B	&   17.926 &   17.285 &   17.282 &   16.517 &   17.268 &   --- &    0.770 &  $105.6\pm 0.4$ & m \\
    1300 	 & 6 58 24.200 & 6 26 52.86 & A	&   18.032 &   17.399 &   17.385 &   16.530 &   17.338 &   --- &    0.817 &  $103.0\pm 1.2$ & m \\
    1302 	 & 6 58 23.298 & 6 24 22.61& A,B	&   17.976 &   17.374 &   17.366 &   16.588 &   17.350 &   --- &    0.791 &  $79.6\pm 0.3$ & n \\
    1327 	 & 6 58 19.564 & 6 26 46.99& A,B	&   18.110 &   17.454 &   17.447 &   16.629 &   17.419 &   --- &    0.749 &  $104.9\pm 0.2$ & m \\
    1337 	 & 6 58 20.259 & 6 22 51.31& A	&   18.130 &   17.500 &   17.490 &   16.666 &   17.474 &   --- &    0.824 &  $105.1\pm 1.0$ & m \\
    1340 	 & 6 57 58.583 & 6 23 05.37& A,B	&   18.143 &   17.473 &   17.463 &   16.599 &   17.487 &   --- &    0.859 &  $110.6\pm 0.2$ & n \\
    1344 	 & 6 58 20.635 & 6 29 20.49& A,B	&   18.210 &   17.538 &   17.527 &   16.652 &   17.496 &   --- &    0.830 &  $54.8\pm 0.5$ & n \\
    1347 	 & 6 58 25.123 & 6 24 50.62& A,B	&   18.113 &   17.559 &   17.545 &   16.769 &   17.514 &   --- &    0.738 &  $-8.4\pm 0.7$ & n \\
    1349 	 & 6 58 06.074 & 6 20 55.42& A,B	&   --- &   --- &   --- &   --- &   17.521 &   --- &    0.834 &  $105.2\pm 1.3$ & m \\
    1350 	 & 6 58 07.071 & 6 29 51.00& A	&   --- &   --- &   --- &   --- &   17.534 &   --- &    0.735 &  $-7.1\pm 0.8$ & n \\
    1358 	 & 6 58 22.953 & 6 26 07.86& A,B	&   18.232 &   17.585 &   17.575 &   16.735 &   17.564 &   --- &    0.790 &  $55.5\pm 1.4$ & n \\
    1359 	 & 6 58 16.363 & 6 20 30.20& A,B	&   --- &   --- &   --- &   --- &   17.565 &   --- &    0.838 &  $112.2\pm 0.7$ & n \\
    1366 	 & 6 57 47.025 & 6 21 04.33& A,B	&   --- &   --- &   --- &   --- &   17.599 &   --- &    0.783 &   $6.9\pm 2.5$ & n \\
    1369 	 & 6 58 00.536 & 6 29 46.04& A,B	&   18.178 &   17.578 &   17.564 &   16.741 &   17.610 &   --- &    0.834 &  $96.8\pm 2.3$ & n \\
    1376 	 & 6 57 56.129 & 6 22 25.39& A,B	&   18.209 &   17.636 &   17.621 &   16.815 &   17.645 &   --- &    0.786 &  $55.2\pm 0.8$ & n \\
    1384 	 & 6 58 20.728 & 6 20 07.11& A,B	&   --- &   --- &   --- &   --- &   17.671 &   --- &    0.871 &  $105.2\pm 0.7$ & m \\
    1396 	 & 6 57 57.167 & 6 22 05.87& A,B	&   --- &   --- &   --- &   --- &   17.729 &   --- &    0.837 &  $84.2\pm 0.7$ & n \\
    1398 	 & 6 57 51.821 & 6 23 45.15& A,B	&   --- &   --- &   --- &   --- &   17.739 &   --- &    0.800 &  $2.2\pm 3.0$ & n \\
    1404 	 & 6 58 16.159 & 6 22 39.07& A,B	&   18.456 &   17.753 &   17.753 &   16.945 &   17.753 &   --- &    0.833 &  $105.2\pm 1.3$ & m \\
    1405 	 & 6 57 58.352 & 6 22 07.42& A,B	&   18.387 &   17.735 &   17.724 &   16.864 &   17.757 &   --- &    0.855 &  $104.6\pm 0.84$ & m \\
    1407 	 & 6 58 00.780 & 6 31 46.58& A,B	&   --- &   --- &   --- &   --- &   17.776 &   --- &    0.891 &  $77.5\pm 0.3$ & n \\
    1410 	 & 6 57 54.497 & 6 20 49.36& A,B	&   --- &   --- &   --- &   --- &   17.788 &   --- &    0.857 &  $53.8\pm 3.7$ & n \\
    1421 	 & 6 58 21.122 & 6 31 28.66& A,B	&   --- &   --- &   --- &   --- &   17.820 &   --- &    0.864 &  $41.3\pm 2.4$ & n \\
    1429 	 & 6 58 23.617 & 6 31 19.37& A,B	&   --- &   --- &   --- &   --- &   17.864 &   --- &    0.833 &  $106.3\pm 1.3$ & m \\
    1433 	 & 6 57 47.116 & 6 24 27.06& A,B	&   --- &   --- &   --- &   --- &   17.890 &   --- &    0.906 &  $61.3\pm 0.9$ & n \\
    1438 	 & 6 58 06.609 & 6 21 07.58& A	&   --- &   --- &   --- &   --- &   17.913 &   --- &    0.874 &  $105.9\pm 1.3$ & m \\
    1448 	 & 6 58 18.537 & 6 21 09.20& A,B	&   --- &   --- &   --- &   --- &   17.943 &   --- &    0.812 &  $56.4\pm 1.4$ & n \\
    1453 	 & 6 58 09.106 & 6 30 41.55& A,B	&   18.671 &   17.977 &   17.959 &   16.986 &   17.955 &   --- &    0.873 &  $104.3\pm 0.6$ & m \\
    1457 	 & 6 58 15.196 & 6 30 33.37& A,B	&   18.691 &   17.966 &   17.947 &   16.939 &   17.962 &   --- &    0.890 &  $73.3\pm 0.5$ & n \\
\end{longtable}
\normalsize
\setcounter{table}{2}
\begin{longtable}{rcccc}
\caption{Final results of our analysis. 
} \label{tab_lith}\\
\hline
ID & (B--V)$_0$& \teff & EW(Li) & \nli\\       
   &           &   (K) & (m\AA) & \\
   &  &  &  &   \\
\hline
\endfirsthead
\caption{continued.}\\
\hline
ID & (B--V)$_0$& \teff & EW(Li) & \nli\\       
   &           &   (K) & (m\AA) & \\
   &  &  &  &   \\
\hline
\endhead
\hline\endfoot
   97 &  0.730 &  5233  &  $40\pm 10$ & $1.53 \pm 0.15$ \\
  109 &  0.522 &  6024  &  $\leq 16 $ &  $\leq 1.88$ \\
  117 &  0.670 &  5401  &  $42\pm 6$  & $1.76 \pm 0.14$ \\
  122 &  0.672 &  5395  &  $\leq 25$  &  $\leq 1.59 $ \\
  130 &  0.490 &  6152  &  $66\pm 7$  & $2.73 \pm 0.12$ \\
  146 &  0.481 &  6189  &  $55\pm 5$  & $2.66 \pm 0.11$ \\
  154 &  0.459 &  6282  &  $58\pm 12$ & $2.77 \pm 0.15$ \\
  193 &  0.454 &  6303  &  $38\pm 7$  & $2.57 \pm 0.15$ \\
  206 &  0.477 &  6206  &  $60\pm 7$  & $2.72 \pm 0.11$ \\
  216 &  0.430 &  6408  &  $63\pm 2$  & $2.92 \pm 0.13$ \\
  231 &  0.466 &  6252  &  $47\pm 2$  & $2.63 \pm 0.13$ \\
  236 &  0.560$^a$ &  5878  &  $48\pm 3$  & $2.31 \pm 0.15$ \\
  241 &  0.481 &  6189  &  $57\pm 4$  & $2.68 \pm 0.14$ \\
  265 &  0.530 &  5992  &  $76\pm 6$  & $2.67 \pm 0.10$ \\
  271 &  0.498 &  6119  &  $59\pm 7$  & $2.64 \pm 0.12$ \\
  276 &  0.504 &  6095  &  $51\pm 9$  & $2.54 \pm 0.13$ \\
  277 &  0.466 &  6252  &  $47\pm 6$  & $2.63 \pm 0.10$ \\
  278 &  0.516 &  6047  &  $57\pm 6$  & $2.56 \pm 0.12$ \\
  288 &  0.583 &  5793  &  $53\pm 4$  & $2.28 \pm 0.10$ \\
  289 &  0.567 &  5852  &  $43\pm 6$  & $2.23 \pm 0.12$ \\
  299 &  0.542 &  5946  &  $65\pm 6$  & $2.54 \pm 0.11$ \\
  311 &  0.546 &  5931  &  $58\pm 6$  & $2.46 \pm 0.12$ \\
  313 &  0.500 &  6111  &  $60\pm 9$  & $2.64 \pm 0.13$ \\
  314 &  0.493 &  6140  &  $59\pm 1$  & $2.66 \pm 0.16$ \\
  315 &  0.546 &  5931  &  $74\pm 7$  & $2.60 \pm 0.11$ \\
  316 &  0.528 &  6000  &  $83\pm 13$ & $2.73 \pm 0.14$ \\
  325 &  0.533 &  5981  &  $63\pm 8$  & $2.55 \pm 0.13$ \\
  333 &  0.540 $^a$&  5954  &  $60\pm 10$ & $2.50 \pm 0.14$ \\
  352 &  0.528 &  6000  &  $70\pm 11$ & $2.63 \pm 0.14$ \\
  357 &  0.551 &  5912  &  $46\pm 7$  & $2.32 \pm 0.17$ \\
  364 &  0.580$^{a,*}$  &  5804  &  $49\pm 7$  & $2.25 \pm 0.14$ \\
  369 &  0.555 &  5897  &  $93\pm 4$  & $2.71 \pm 0.15$ \\
  371 &  0.571 &  5837  &  $84\pm 12$ & $2.59 \pm 0.14$ \\
  374 &  0.583 &  5793  &  $50\pm 10$ & $2.25 \pm 0.15$ \\
 1050 &  0.853 &  4920  &  $65\pm 5$  & $1.45 \pm 0.16$ \\
 1070 &  0.522 &  6024  &  $25\pm 7$  & $2.12 \pm 0.25$ \\
 1092 &  0.473 &  6222  &  $33\pm 5$  & $2.44 \pm 0.15$ \\
 1113 &  0.472 &  6227  &  $38\pm 3$  & $2.51 \pm 0.15$ \\
 1121 &  0.753 &  5233  &  $\leq 17$  & $\leq 1.82$ \\
 1128 &  0.450$^b$ &  6320  &  $31\pm 6$  & $2.49 \pm 0.14$ \\
 1130 &  0.450$^b$ &  6320  &  $\leq 20$  & $\leq 2.27$ \\
 1150 &  0.480$^b$ &  6193  &  $55\pm 5$  & $2.56 \pm 0.12$ \\
 1212 &  0.475 &  6214  &  $55\pm 7$  & $2.68 \pm 0.12$ \\
 1228 &  0.465 &  6256  &  $63\pm 4$  & $2.79 \pm 0.12$ \\
 1231 &  0.457 &  6290  &  $56\pm 6$  & $2.76 \pm 0.11$ \\
 1241 &  0.530$^{b,*}$  &  5992  &  $52\pm 5$  & $2.46 \pm 0.11$ \\
 1257 &  0.490$^b$ &  6152  &  $53\pm 9$  & $2.61 \pm 0.12$ \\
 1263 &  0.490$^b$ &  6152  &  $60\pm 7$  & $2.68 \pm 0.12$ \\
 1279 &  0.501 &  6107  &  $61\pm 10$ & $2.65 \pm 0.13$ \\
 1300 &  0.493 &  6140  &  $42\pm 6$  & $2.48 \pm 0.12$ \\
 1327 &  0.516 &  6047  &  $37\pm 10$ & $2.34 \pm 0.19$ \\
 1337 &  0.490 &  6152  &  $44\pm 5$  & $2.51 \pm 0.12$ \\
 1349 &  0.530 $^b$&  5992  &  $85\pm 9$  & $2.74 \pm 0.12$ \\
 1384 &  0.570 $^b$&  5841  &  $59\pm 6$  & $2.39 \pm 0.12$ \\
 1405 &  0.512 &  6063  &  $65\pm 10$ & $2.64 \pm 0.14$ \\
 1438 &  0.570 $^b$&  5841  &  $70\pm 12$ & $2.48 \pm 0.15$ \\
 1453 &  0.554 &  5900  &  $49\pm 9$ & $2.34 \pm 0.15$ \\
      &       &        &            &                 \\
\multicolumn{5}{l}{$a$: B--V from Richtler \& Sagar (\cite{rs00})}\\
\multicolumn{5}{l}{$b$: (B--V)$_0$ derived from (V--I) (see text)}\\
\multicolumn{5}{l}{$*$: B--V from D'Orazi et al. available, but not
used (see text)}\\
\hline
\end{longtable}
\end{document}